\documentclass[aps,prc,reprint,superscriptaddress,showpacs,twocolumn]{revtex4}

\usepackage{natbib}
\usepackage{graphicx}
\usepackage{textcomp}
\usepackage{multirow}
\begin{document}



\title{Isotopic distribution of fission fragments in collisions between $^{238}$U beam and $^9$Be and $^{12}$C targets at  24 MeV/u}


\author{O.~Delaune}
\email[]{olivier.delaune@cea.fr}
\altaffiliation[Present address:]{CEA DAM DIF, F-91297 Arpajon, France }
\affiliation{Grand Acc\'el\'erateur National d'Ions Lourds, CEA/DSM-CNRS/IN2P3, F-14076 Caen, France}
\author{F.~Farget}
\email[]{fanny.farget@ganil.fr}
\affiliation{Grand Acc\'el\'erateur National d'Ions Lourds, CEA/DSM-CNRS/IN2P3, F-14076 Caen, France}
\author{O. B. Tarasov}
\affiliation{NSCL, Michigan State University, East Lansing, MI 48824, USA}
\author{A. M. Amthor}
\altaffiliation[Present address:]{NSCL, Michigan State University, East Lansing, MI 48824, USA}
\affiliation{Grand Acc\'el\'erateur National d'Ions Lourds, CEA/DSM-CNRS/IN2P3, F-14076 Caen, France}
\author{B. Bastin}
\affiliation{Grand Acc\'el\'erateur National d'Ions Lourds, CEA/DSM-CNRS/IN2P3, F-14076 Caen, France}
\author{D. Bazin}
\affiliation{NSCL, Michigan State University, East Lansing, MI 48824, USA}
\author{B. Blank}
\affiliation{CENBG, UMR 5797 CNRS/IN2P3, Universit\'e Bordeaux 1, F-33175 Gradignan, France}
\author{L.~Cac\'eres}
\affiliation{Grand Acc\'el\'erateur National d'Ions Lourds, CEA/DSM-CNRS/IN2P3, F-14076 Caen, France}
\author{A.~Chbihi}
\affiliation{Grand Acc\'el\'erateur National d'Ions Lourds, CEA/DSM-CNRS/IN2P3, F-14076 Caen, France}
\author{B.~Fern\'andez-Dom\'inguez}
\affiliation{Universidade de Santiago de Compostela, E-15782 Santiago de Compostela, Spain}
\author{S. Gr\'evy}
\affiliation{CENBG, UMR 5797 CNRS/IN2P3, Universit\'e Bordeaux 1, F-33175 Gradignan, France}
\author{O. Kamalou}
\affiliation{Grand Acc\'el\'erateur National d'Ions Lourds, CEA/DSM-CNRS/IN2P3, F-14076 Caen, France}
\author{S.~Lukyanov}
\affiliation{FLNR, JINR, 141980 Dubna, Moscow region, Russian Federation}
\author{W. Mittig}
\affiliation{NSCL, Michigan State University, East Lansing, MI 48824, USA}
\affiliation{Dep. of Physics and Astronomy, Michigan State University, East 
Lansing, MI 48824, USA}
\author{D.~J.~Morrissey}
\affiliation{NSCL, Michigan State University, East Lansing, MI 48824, USA}
\affiliation{Dep. of Chemistry, Michigan State University, East Lansing, MI 48824, USA}
\author{J.~Pereira}
\affiliation{NSCL, Michigan State University, East Lansing, MI 48824, USA}
\author{L.~Perrot}
\affiliation{IPN Orsay, CNRS/IN2P3, F-91406 Orsay, France}
\author{M.-G.~Saint-Laurent}
\affiliation{Grand Acc\'el\'erateur National d'Ions Lourds, CEA/DSM-CNRS/IN2P3, F-14076 Caen, France}
\author{H.~Savajols}
\affiliation{Grand Acc\'el\'erateur National d'Ions Lourds, CEA/DSM-CNRS/IN2P3, F-14076 Caen, France}
\author{B. M. Sherrill}
\affiliation{NSCL, Michigan State University, East Lansing, MI 48824, USA}
\affiliation{Dep. of Physics and Astronomy, Michigan State University, East Lansing, MI 48824, USA}
\author{C. Stodel}
\affiliation{Grand Acc\'el\'erateur National d'Ions Lourds, CEA/DSM-CNRS/IN2P3, F-14076 Caen, France}
\author{J. C. Thomas}
\affiliation{Grand Acc\'el\'erateur National d'Ions Lourds, CEA/DSM-CNRS/IN2P3, F-14076 Caen, France}
\author{A. C. Villari}
\affiliation{Grand Acc\'el\'erateur National d'Ions Lourds, CEA/DSM-CNRS/IN2P3, F-14076 Caen, France}

\date{\today}

\begin{abstract}
Inverse kinematics coupled to a high-resolution spectrometer is used to investigate the isotopic yields of fission fragments produced in reactions between a $^{238}$U beam at 24 MeV/u and $^9$Be and $^{12}$C targets. Mass, atomic number and isotopic distributions are reported for the two reactions. These informations give access to the neutron excess and the isotopic distribution widths, which together with the atomic-number and mass distributions are used to investigate the fusion-fission dynamics. 
\end{abstract}


\maketitle


\section{\label{sect:intro}Introduction}

In heavy-ion induced fission reactions, the fission-fragment production results from a complicated process where the collision stage of the reaction influences greatly the formation of the compound nucleus, which then deexcites in converting a part of its excitation energy into deformation energy, up to the two fragments inception and their final separation at the scission elongation. The evolution of the nucleus binding energy with deformation defines a potential energy landscape. It is shaped by the macroscopic properties of the nucleus, and may be strongly influenced by the single-particle structure of the nucleus if the excitation energy is sufficiently low. The fission-fragment distribution is directly linked to the deformation path, which may be affected by dissipative effects. In nuclear reactions leading to compound nuclei with excitation energy sufficient to minimize significantly the structural effect, the fission fragments are sensitive to the evolution of the macroscopic part of the potential energy, and to the dissipation that slows down the deformation process, inducing a competition between neutron evaporation and deformation. Fragment mass distributions in fusion-fission reactions have been investigated intensively~\cite{ThH08,ItB07}, as they are a witness of the nuclear reaction all along its process, from the collision stage, that may lead to quasi-fission, to the formation conditions of the compound nucleus, which finally deforms up to the scission point, and the subsequent neutron evaporation by the separated fragments.  These investigations are also of importance as the reaction dynamics determine the formation probability of superheavy elements in fusion reactions~\cite{Zag04}. To understand in more details the fission process, measurement of the isotopic distributions of fission fragments have been made using $\gamma$-ray detection emitted by the fragments~\cite{BoK07}. However, this technique suffers from inappropriate life-times in specific nuclei, or incomplete information on the decay path, inducing large fluctuations in the resulting yields. 

The present work describes a novel method, which allows to get in addition to the fission-fragment mass distributions, the isotopic fission-fragment yields over the entire atomic-number range of the fission fragments (from $Z$=31 to $Z$=64), using inverse kinematics coupled to the LISE spectrometer~\cite{AnB87}. 
The fusion-fission reactions are induced using a $^{238}$U beam on light targets of $^9$Be and $^{12}$C. The substantial asymmetry of the reaction hinders dissipative effects in the collision stage of the reaction, and therefore quasi-fission is 
negligible~\cite{Swi81}. The resulting data give insight into the compound nucleus formation and the deexcitation stage of the reaction. In particular, the neutron excess of the fission-fragments  reflects the proton-to-neutron equilibration during deformation, followed by the sharing of the excitation energy by the two fragments.

In addition to their interest for nuclear-reaction dynamics, fusion-fission fragment yields are an interesting probe to test the technique to produce in-flight exotic-ion beams in inverse kinematics asymmetric fusion-fission reactions. Indeed, in these reactions the fusion cross section is large, and the weak slowing down of the beam in the light atomic-number target gives the possibility to use thick targets. As a result high yields of exotic nuclei are expected. In addition, high-energy fission allows to produce nuclei in the valley region, as well as on a broader range of elements~\cite{TaV08}. 

GANIL delivers $^{238}$U beam accelerated up to 24 A MeV. This energy has been chosen to bombard thick $^9$Be and $^{12}$C targets, leading to compound nuclei of $^{247}$Cm and $^{250}$Cf. Section~\ref{sect:exp} describes the experiment and the fission-fragment identification, section~\ref{sect:rec} describes the procedure to obtain fission yields in both reactions, and the resulting isotopic yields of these two systems are presented in section~\ref{sect:res}.

\section{\label{sect:exp}Experiment}

\begin{figure}[t]
\hspace{0cm}\includegraphics[width=1\linewidth]{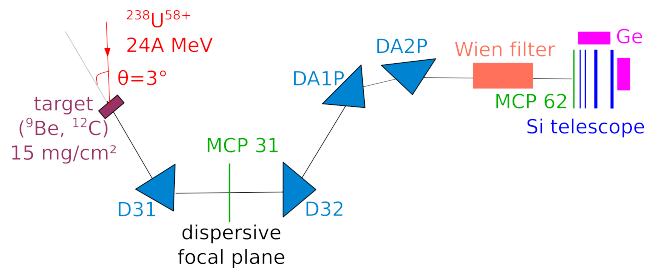}
 \caption{\label{setup}Scheme of the LISE spectrometer with its detector equipment used for the identification of the fission fragments.}
 \end{figure}

The $^{238}$U$^{58+}$ beam has been accelerated up to 24 MeV/u with the two cyclotrons CSS1 and CSS2 of GANIL. It impinged the LISE target station with an angle of 3 $^\circ$ to prevent the beam that did not interact with the target to enter the spectrometer, as depicted in figure~\ref{setup}. Two targets  made of $^9$Be and $^{12}$C material have been irradiated, both were 15 mg/cm$^2$ thick.  The beam energy-loss in the target is 1.222 and 1.337 GeV for Be and C target, respectively. As summarised in table~\ref{table1}, in the reference frame of the centre of mass the reaction takes place with an average energy of  187 and 243 MeV for the Be and the C target, the Coulomb barrier between beam and target nuclei being of 43.4 and 64.5 MeV, respectively. 

\begin{center}
\begin{table}
  \begin{tabular}{@{}| l|c|c| @{}}
    \hline
    Target & $^9$Be & $^{12}$C \\ 
    Available energy in the centre of mass (MeV)& 187 & 243 \\ 
    Coulomb barrier (MeV)& 43.4 & 64.5 \\ 
    Angular momentum ($\hbar$)& 57 & 80 \\ 
    \hline
  \end{tabular}
\caption{\label{table1} Entrance channel characteristics for the two reactions considered. The available energy in the centre of mass is calculated at the middle of the target thickness and corresponds to the average energy.}
\end{table}
\end{center}

\begin{figure}[t]
\includegraphics[width=1\linewidth]{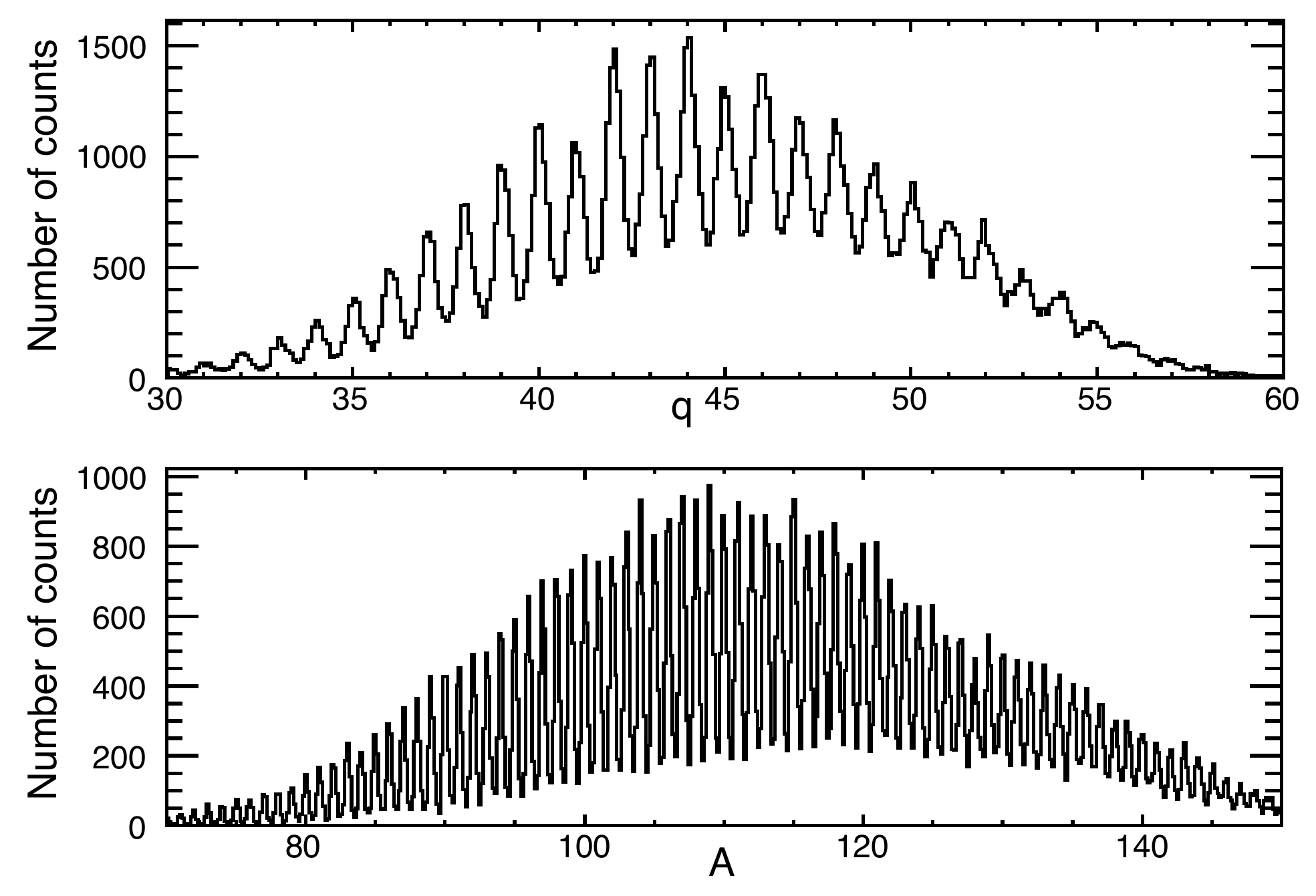}
 \caption{\label{dist_qa}Upper panel: Ionic charge-state distribution measured for the nominal magnetic rigidity $B\rho_0$ = 1.9 Tm. Lower panel : Mass distribution for the same spectrometer setting.}
 \end{figure}

Considering the high fission probability of the excited heavy nuclei produced in the collision, most of the reactions lead to the emission of fission fragments that  are emitted in a cone of about 10$^\circ$ of aperture. The tiny proportion that enters the LISE spectrometer~\cite{AnB87} is then identified based on  magnetic-rigidity $B\rho$, time-of-flight $ToF$, energy $E$ and energy-loss $\Delta E$ measurements. Two position-sensitive micro-channel plate detectors~\cite{OdW96} measured the position of the particles $X_{31}$ and $X_{62}$  at the intermediate dispersive plane and the final focal plane, respectively, to deduce the magnetic rigidity of the particles, following the equation:
\begin{equation}
B\rho=B\rho_0(1+\frac{X_{62}}{D_{62}}-\frac{MX_{31}}{D_{62}}),
\end{equation}
where $B\rho_0$ is the nominal value of the spectrometer magnetic rigidity, $M$ and $D_{62}$ are respectively the magnification and the dispersion in the second section of the spectrometer after the intermediate focal plane.  During the experiment, the spectrometer sections before and after the intermediate focal plane were set at an identical value of nominal magnetic rigidity. The position calibration of the micro-channel detectors was performed using slits placed in front of the detector. A series of slit-aperture positions let the fission-fragment beam irradiating the detectors at different calibrated positions. The spectrometer parameters (dispersions and magnification) have been calibrated with position measurements of different charge states of the beam.  
Downstream the second 
micro-channel plate detector, a stack of four silicon detectors was installed to measure the energy loss and the energy of the ions. Two germanium detectors were installed around the silicon-detector stack to detect eventual gamma rays emitted by the stopped fragments.  The experimental set-up is outlined in figure~\ref{setup}. 
The time-of-flight measurement was processed between the position-sensitive micro-channel plate detector placed at the intermediate focal plane of the spectrometer, and the first silicon detector placed at the final focal plane. The flight path was considered independent of the position measured and equals 32.423 m. The time-of-flight was calibrated with the direct beam passing through the spectrometer. The measurement of the magnetic rigidity $B\rho$ and the velocity $v$ gave access to the ratio $\frac{A}{q}$ of the fragments, where $A$ is their mass number and $q$ their ionic charge-state from the relation:
\begin{equation}
B\rho=3.107\frac{A}{q}\beta\gamma,
\end{equation}where $\beta = \frac{v}{c}$ is the reduced velocity, $c$ being the velocity of light, $\gamma$ is the associated Lorentz factor, and $B\rho$ is expressed in Tm.
The ionic charge-state of the fragments was deduced from the combined measurements of their magnetic rigidity and of their kinetic energy $KE$ in the stack of silicon detectors. The energy of the fragments is related to the mass of the fragments $A_E$ and their velocity:
\begin{equation}
KE=m_0A_E(\gamma-1),
\end{equation}
where $m_0$ is the atomic mass unit. The ionic charge-state $q$ is then deduced from the ratio between $A_E$ and $\frac{A}{q}$. The ionic charge-state distribution measured with the nominal magnetic rigidity of 1.9 Tm is displayed in the upper panel of figure~\ref{dist_qa} . 
The resolution obtained $\frac{\Delta q}{q} = 2 \%$ is governed by the silicon-detector energy resolution. Once the ionic charge state is determined, its integer value is used to factorize the ratio $\frac{A}{q}$ to deduce the mass of the fragments. The corresponding mass distribution is displayed in the  lower panel of figure~\ref{dist_qa}. A resolution of $\frac{\Delta A}{A} = 0.5\% $ FWHM was achieved, which gave a good separation over the complete mass distribution. 

\begin{figure}[t]
\includegraphics[width=1\linewidth]{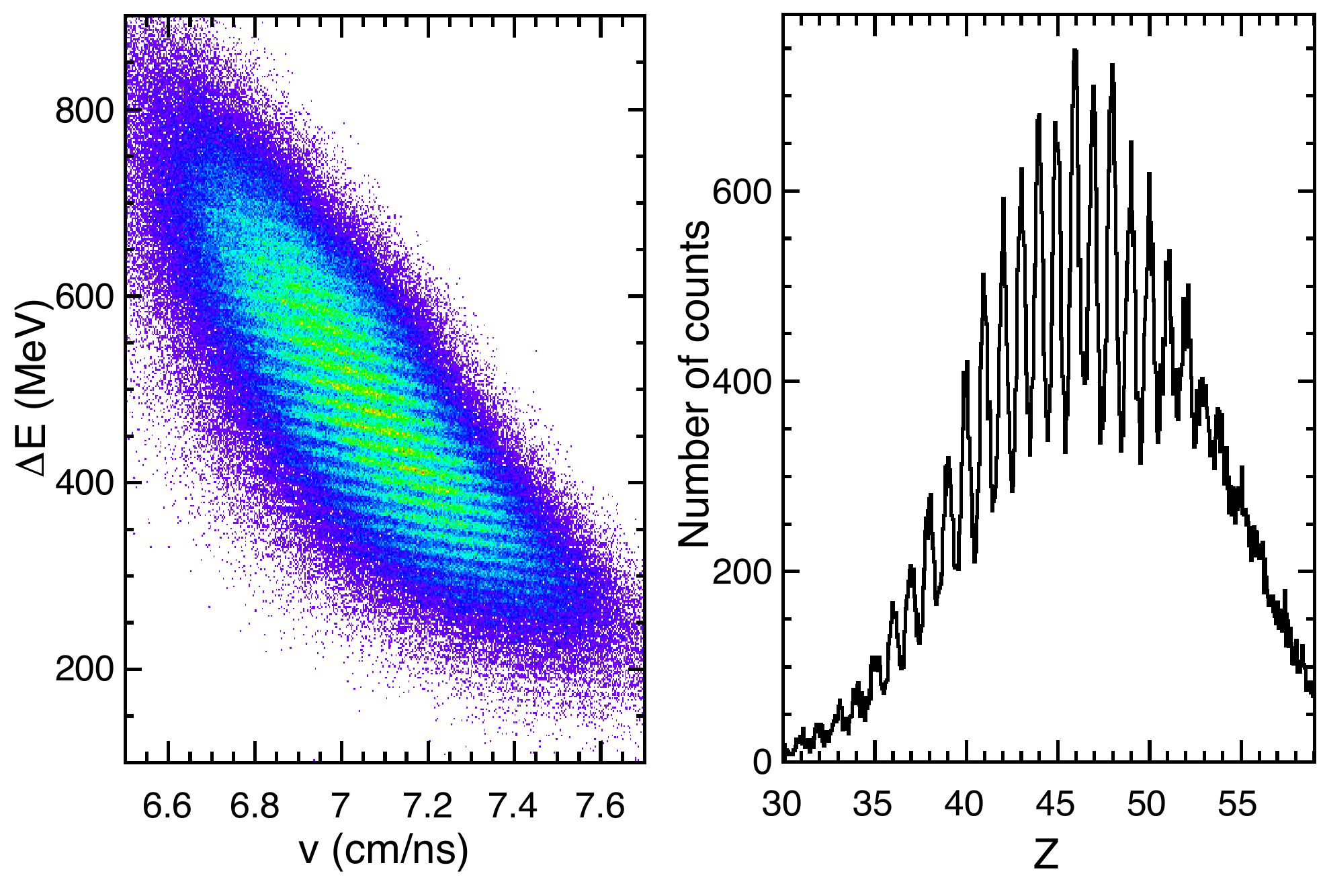}
 \caption{\label{dist_Z}Left panel: Energy-loss $\Delta E$ displayed as a function of the fragment velocity v for the nominal magnetic rigidity 1.9 Tm. Right panel : Corresponding atomic number distribution.}
 \end{figure}

The atomic number $Z$ of the fragments was identified with the energy-loss measurement in the first silicon detector of the silicon stack, with a thickness of 69 $\mu$m. The energy-loss of the fragments as a function of their velocity measured for the nominal magnetic rigidity $B\rho_0 = 1.9$ Tm is displayed in the left panel of figure~\ref{dist_Z}. The energy-loss was corrected for its velocity dependency, following a fit of the different energy-loss lines by first order polynomial functions. The corresponding atomic number distribution is displayed in the right panel of figure~\ref{dist_Z}. 

\begin{figure}[t]
\includegraphics[width=0.7\linewidth]{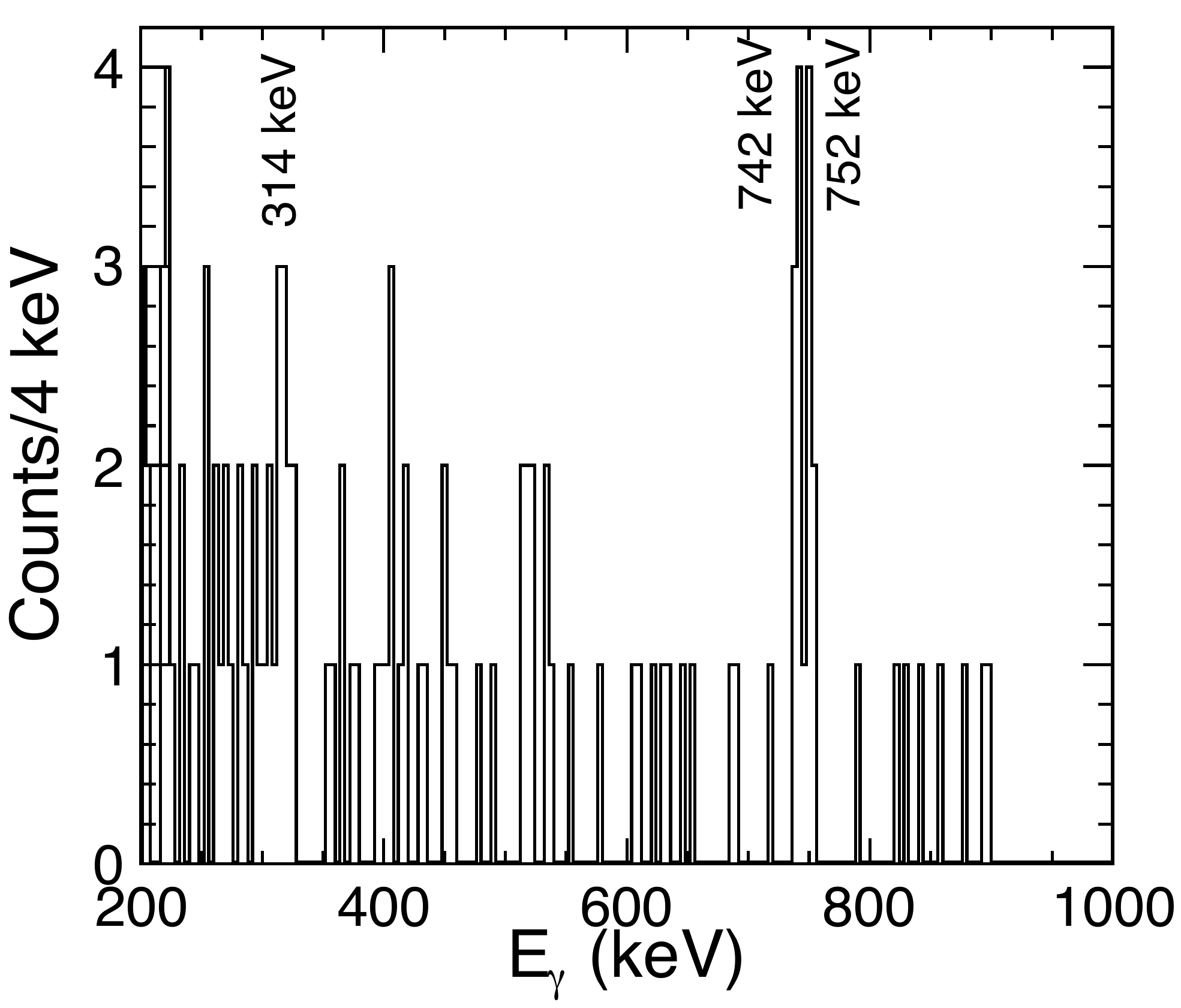}
 \caption{\label{gamma} Gamma-ray spectrum observed in coincidence with $^{128}Te$. The characteristic gamma lines of 314, 742 and 752 keV sign the decay of the isomeric state of $\tau_{\frac{1}{2}}$ = 370 ns known in this nucleus.}
 \end{figure}

The identification in atomic number and mass number was strengthened by an iterative procedure, where for one supposed identified fragment, the measured energy loss, velocity and residual energy was checked to match with simulations of these quantities with the LISE++ software~\cite{LISE++}. A final confirmation of the identification was performed by the observation of the decay of isomeric states populated in fission fragments with two germanium detectors placed around the silicon stack where the fission fragments stopped. Due to the limited beam-time, low statistics spectra were obtained, but as can be seen in figure~\ref{gamma}, the isomeric state of half-life $\tau_{\frac{1}{2}}$ = 370 ns, decaying after the flight in the spectrometer, confirmed the identification of the  $^{128}Te$ nucleus. 

\subsection{\label{sect:qs}Beam ionic charge-state distribution}

\begin{figure}[h]
\includegraphics[width=1\linewidth]{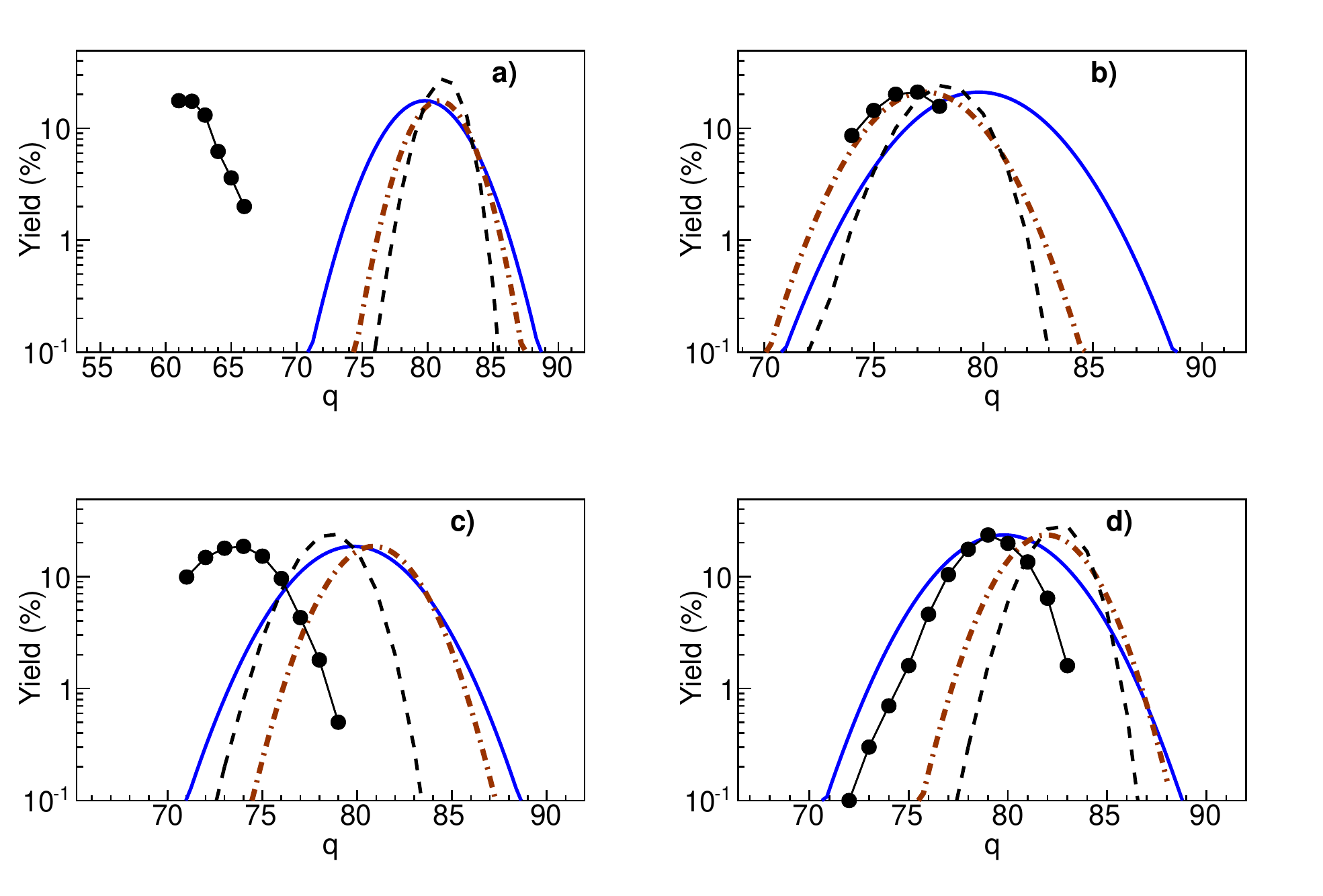}
 \caption{\label{bqstate} Primary-beam ionic charge-state distributions measured behind different materials. a) 40$\mu$g/cm$^2$ C; b) 3mg/cm$^2$ Al, c)15$\mu$g/cm$^2$ Mylar foil and 20$\mu$g/cm$^2$ Al; d) 1.5mg/cm$^2$ Be. They are compared to different parameterizations for charge-state distributions: Schiwietz and Grande~\cite{ScG01} (solid blue line), 
 Leon~\cite{LeM98}  (dashed red line), Winger~\cite{WiS92}(black dashed line). }
 \end{figure}

The spectrometer dispersion of 1.8 cm/\% gives the possibility to detect several ionic charge-states of the uranium beam within one unique magnetic-field setting, and an absolute calibration of the spectrometer characteristics. Different values of the spectrometer nominal magnetic rigidity where scanned in order to cover the complete ionic charge-state distribution of the beam. The resulting charge-state distributions are displayed in figure ~\ref{bqstate}.   They are compared to different parameterizations typically used at this energy. Left panels in figure~\ref{bqstate} show the beam ionic charge-state distributions after thin Carbon and Mylar layers. It is clear that the thin layers barely strip the incoming beam with ionic charge-state of 58. Consequently, the different parameterizations show an important discrepancy with the experimental results as they consider the ionic charge-state distributions after a layer sufficient to reach the equilibrium charge-state, which is not the case with such thin layers.  In the right panels, layers thick enough to reach the equilibrium charge-state have been used, and the experimental data show a stronger stripping effect, the average charge-state being increased from 58 in front of the target to 76 and 79 behind the Al and Be target, respectively. A much better agreement is observed with the different parameterizations. The Leon prescription~\cite{LeM98} gives excellent results after the thick Al layer, whereas it is significantly too high in the case of Be target. In both cases, the Schiwietz and Grande parameterization~\cite{ScG01}  gives a fair prediction of the average charge state, while the width of the distribution is too wide. 

\section{\label{sect:rec}Reconstruction of the fission-fragment yields}
 
\begin{figure}[h]
\includegraphics[width=1\linewidth]{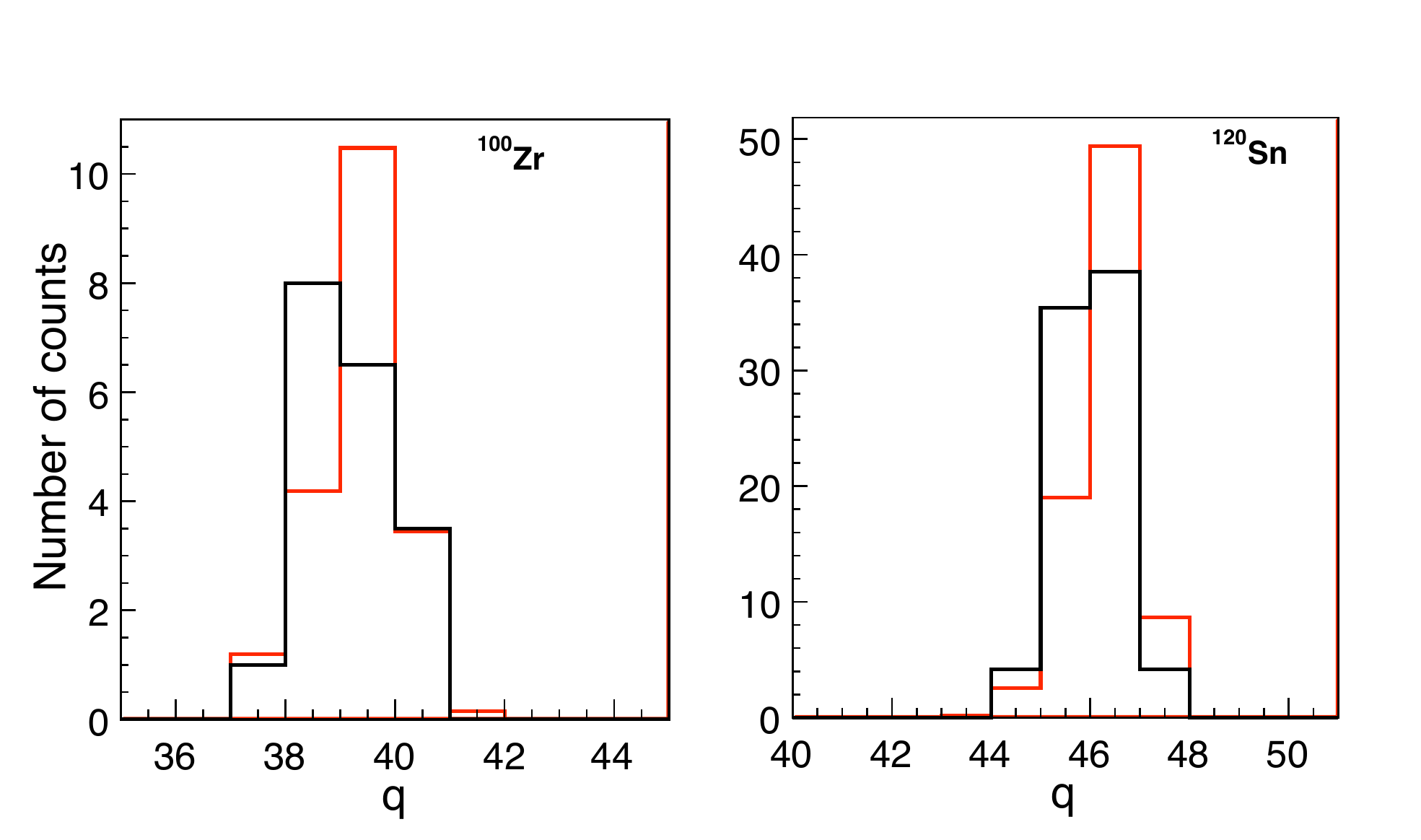}
 \caption{\label{qstate} Ionic charge state distributions measured for $^{100}$Zr and $^{120}$Sn in red, compared to the results of the simulation (see text for details), in black.}
 \end{figure}

The spectrometer induced cuts in the angular distribution of the fission fragments, as well as in their momentum distribution. The angle-aperture is $\pm 1^\circ$, and the magnetic rigidity acceptance $\pm 0.8\%$ around the nominal magnetic rigidity value. The fragment production was measured for several values of the nominal magnetic rigidity in order to cover the fragment momentum distribution. Due to the limited beam-time, only four different values of the nominal rigidities could be measured, and in order to cover as much as possible of the fragment momentum distribution, the nominal values were separated by about 5 \%. To estimate the losses, a simulation of the kinematics of the fission fragments was performed, based on the assumption that in the reference frame of the fissioning system, the total kinetic energy $TKE$ is shared between the two fragments:
\begin{equation}
TKE = \frac{1}{2}A_h{v_h}^2+\frac{1}{2}A_l{v_l}^2,
\end{equation}
where the indexes $_h$ and $_l$ refer to the heavy and light fragment, respectively, and that the total kinetic energy can be expressed in MeV following~\cite{WiS76}:
\begin{equation}
TKE=1.44\frac{Z_hZ_l}{r_0(A_h^{1/3}(1+\frac{2}{3}\beta_h)+A_l^{1/3}(1+\frac{2}{3}\beta_l)) + d}.
\end{equation}
The radius of the nucleon $r_0$ was assumed to be 1.16 fm, the parameters $\beta_h$ and $\beta_l$, referring to the deformation of the heavy and light fission-fragments at the scission point have been taken equal to 0.625, and the tip distance between the two fragments $d$ equal to 2 fm, as proposed in~\cite{WiS76} and confirmed in~\cite{BoS97}. For each fragment, the total kinetic energy was calculated assuming that $A_h+A_l=A_f$, and $Z_h+Z_l=Z_f$, where $A_f$ and $Z_f$ are the mass and the atomic number of the fissioning nucleus. The kinematic characteristics of each fragment were then transferred into the laboratory reference frame, assuming a random position along the target thickness for the reaction to take place, inducing a wide distribution of the fissioning-nucleus velocity. The ionic charge-state distribution of the fragments was estimated according to the Schiwietz and Grande parameterization~\cite{ScG01}. To reproduce the data a  scaling factor of 1.025 was applied to the mean charge-state, while the width of the ionic charge-state distribution was not modified. Finally, the angular cuts of the spectrometer were applied to the kinematics simulation. Figure~\ref{qstate} shows a comparison between the results of the simulation for different ions, and the experimental charge-state distribution measured for the ensemble of the spectrometer settings during the experiment. The good agreement between the simulated and measured ionic charge-state distribution gives confidence in the correct simulation of the kinematics and the charge state distribution.  

\begin{figure}[t]
\includegraphics[width=0.8\linewidth]{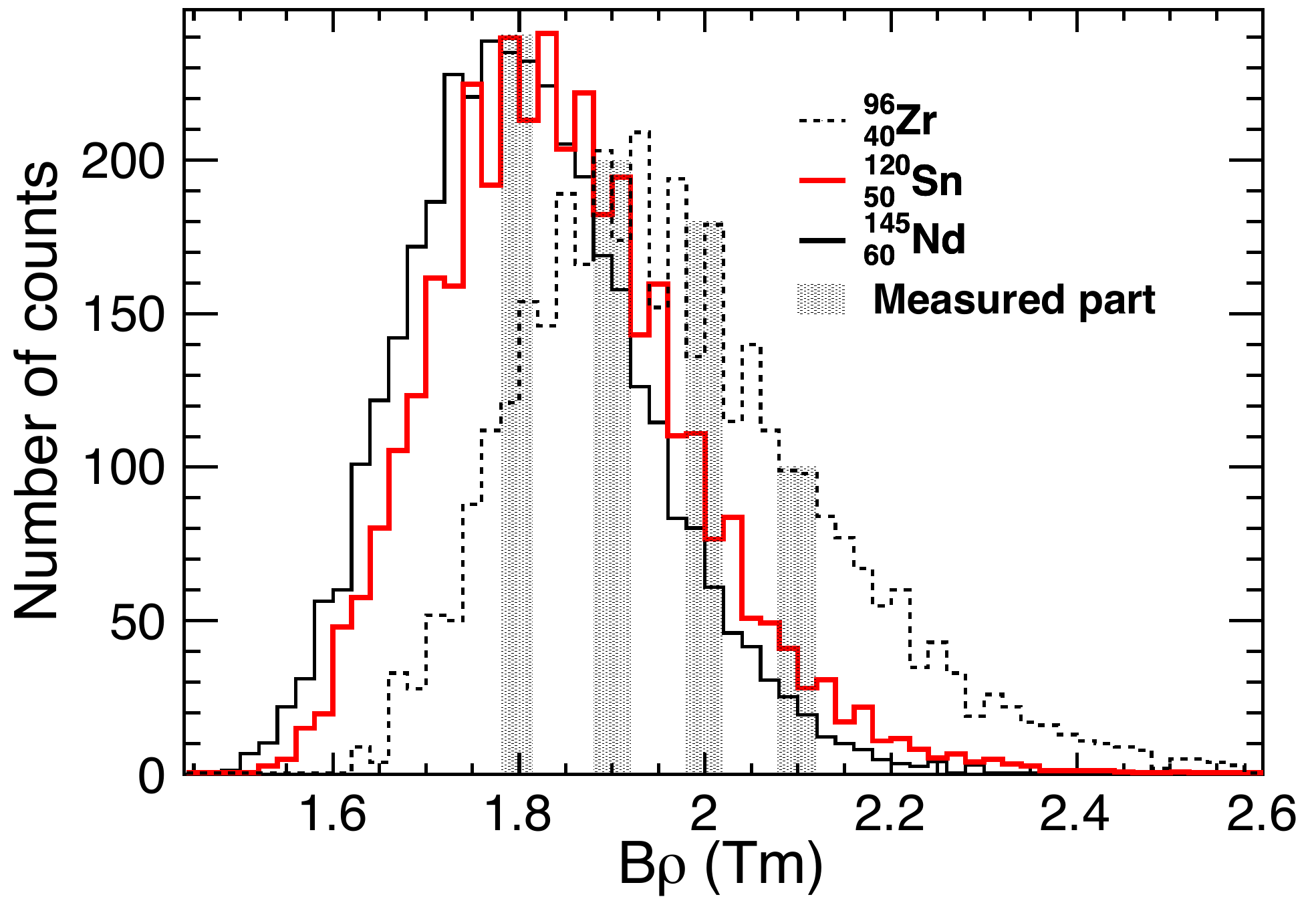}
 \caption{\label{kin} simulation of the fission-fragment magnetic rigidity distribution for different isotopes: $^{96}$Kr (dashed line),  $^{120}$Sn (grey full line, red color online), and $^{145}$Nd (black line). The angular trnasmission of the spectrometer of 3$\pm 1^\circ$ has been applied, and only forward fission is considered. The magnetic rigidity acceptance of $\pm 0.8\%$ for the 4 spectrometer settings of 1.8, 1.9, 2.0 and 2.1 Tm are shown as shaded area.}
 \end{figure}
\begin{figure}[t]
\includegraphics[width=1\linewidth]{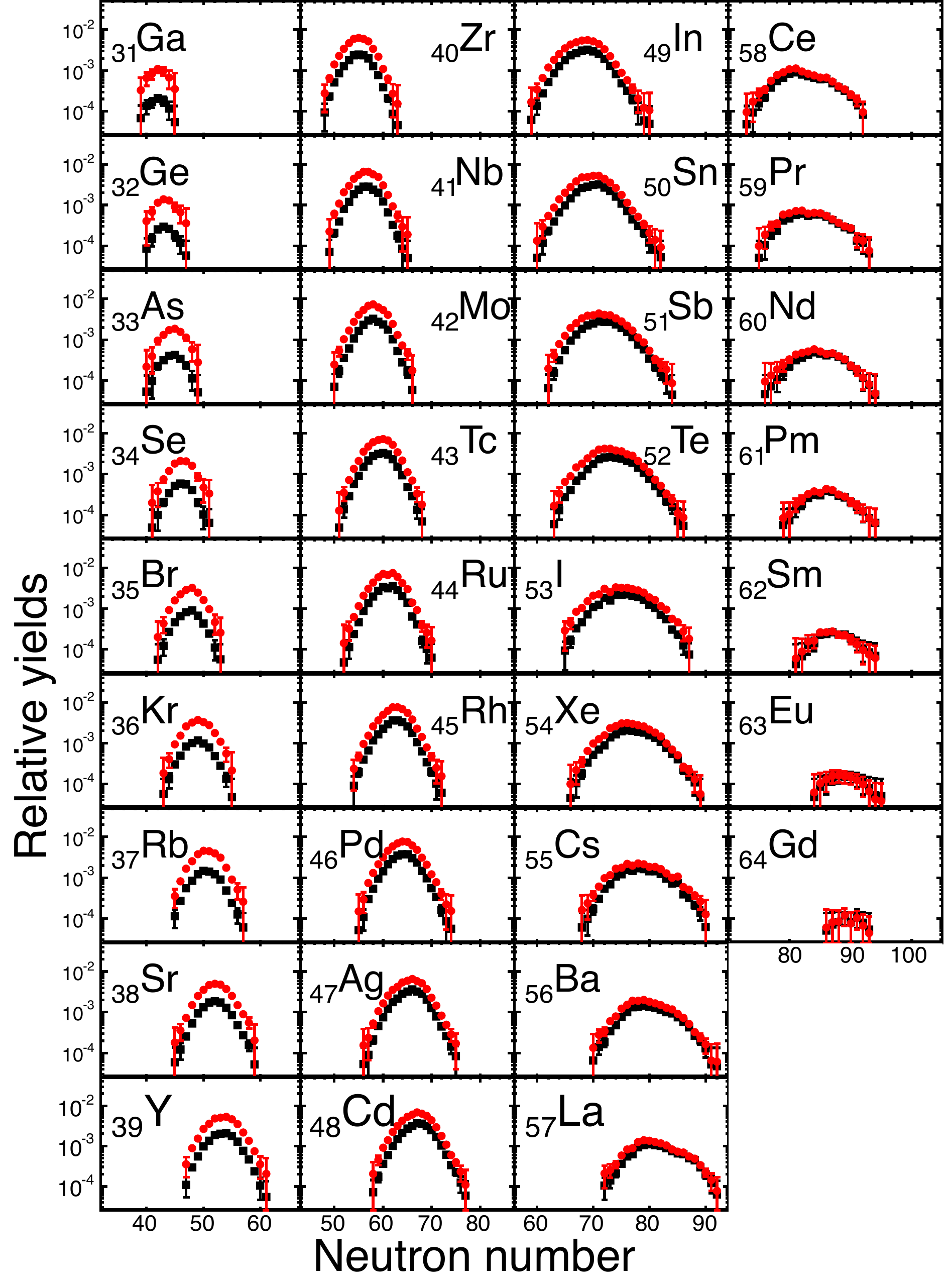}
 \caption{\label{yields_corr} Fission fragment isotopic yields measured in the reaction $^{238}$U+$^{12}$C. Raw data and data corrected for the LISE spectrometer acceptance losses are shown in black and red, respectively.}
 \end{figure}

The simulated magnetic-rigidity distributions for different fission fragments are displayed in figure ~\ref{kin}, where the angular transmission of 3$\pm 1^\circ$ has been applied. Only fission fragments emitted in the direction of the beam (forward fission) are considered in the figure. For each isotope, the width of the magnetic rigidity distribution is the convolution of  the ionic charge-state distribution and the velocity distribution. The velocity-distribution width results from the fission kinematics convoluted with the large energy straggling in the target, which is the most important factor in the magnetic rigidity spread.  
In the simulation, a flat random distribution in the atomic and mass numbers of the fission fragments was used, and for each isotope the correction factor $f_{acc}(Z,A)$ for the angular and magnetic-rigidity cuts was defined as the ratio of  the number of fragments generated and the number of fragments transmitted in the angular acceptance and the sum of the four different spectrometer settings:  $f_{acc} (Z,A) =N_{produced}(Z,A)/N_{transmitted}$(Z,A). In figure ~\ref{kin}, it corresponds to the ratio between the summed shaded area, and the total area of the magnetic rigidity spectrum. 
The yields measured for the four different spectrometer settings were normalized to the incident beam intensity, which was measured at the start and end of each run with a Faraday cup inserted at the target station. For each fragment, the number of counts  normalized to the beam intensity was then multiplied by the correction factor $f_{acc}(Z,A)$.  As can be seen in figure~\ref{yields_corr}, the correction factor did not modify substantially the measured isotopic distributions. Indeed, despite the incomplete coverage of the magnetic rigidity distributions during the experiment, the four different spectrometer settings span over most of the magnetic rigidity distribution, as sketched in figure~\ref{kin}, ensuring for a good estimation of the produced yields. Due to the stronger focussing of the heavy fragments, their correction factor is smaller than for the light fragments. Finally, the yields were normalized such that the sum over the entire distribution is equal to 200.

 \section{\label{sect:res}Results and discussion}
 
\begin{figure}[t]
 \includegraphics[width=1\linewidth]{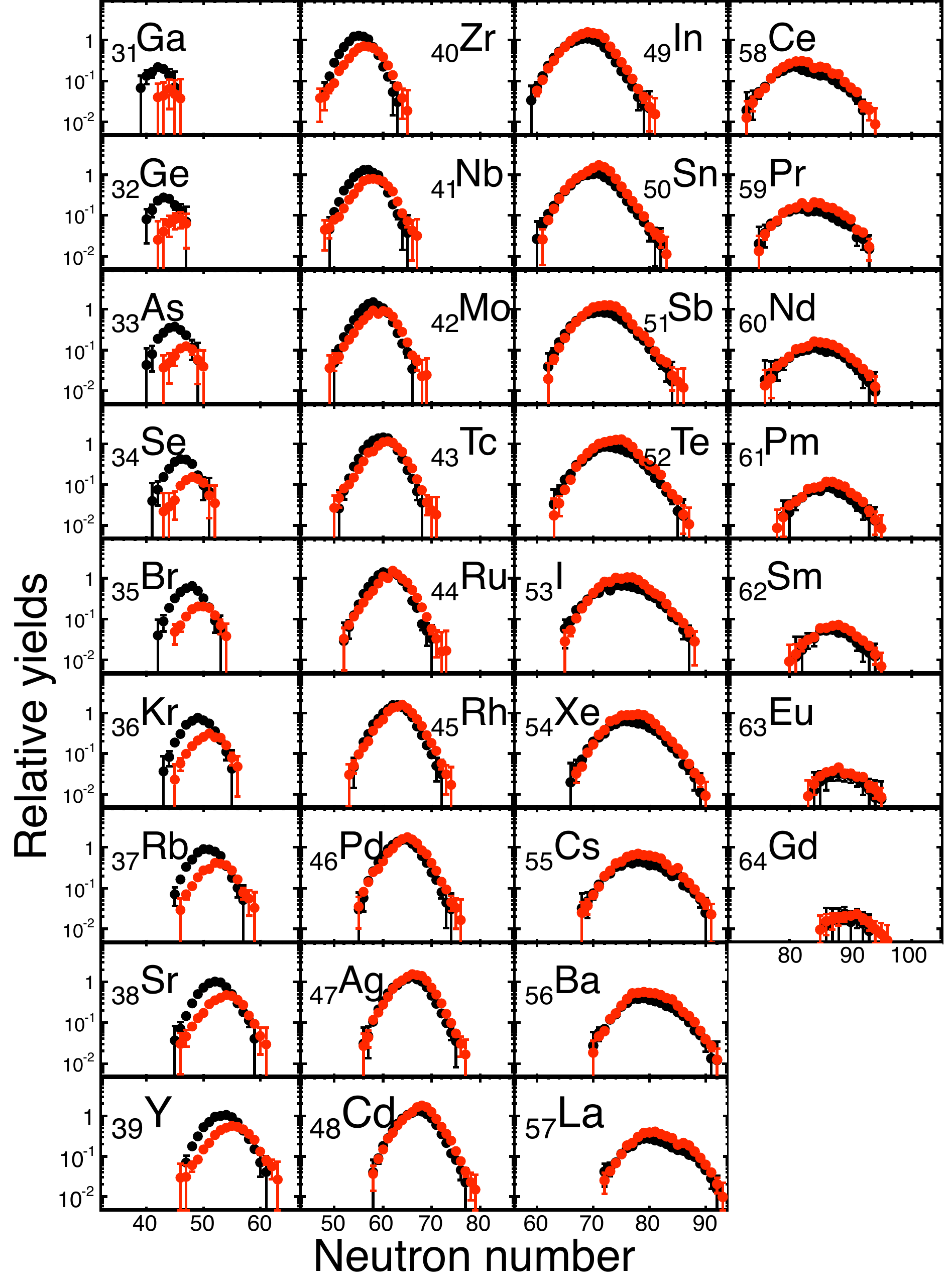}
\caption{\label{yields} Isotopic fission-fragment yields measured in the reaction $^{238}$U+$^9$Be (red), and $^{238}$U+$^{12}$C (black). }
 \end{figure}

Isotopic-yield distributions from Ga to Gd, obtained with the method described in the preceding section, for the two reactions considered in this work, are displayed in figure~\ref{yields}. A typical bell shape for the fission-fragment distribution is observed in both reactions. The distributions for the heavy elements ($Z>45$) show a good superimposition, with substantially higher yields of neutron-rich nuclei in the $^{238}$U+$^{9}$Be reaction;  neutron-rich fragments yields are produced with a factor of 50\% to 100\% more than in the reaction induced on $^{12}$C. A systematic shift towards more neutron-deficient isotopes of the fragment distributions produced in the $^{238}$U+$^{12}$C reaction is observed as the atomic number of the fragments decreases. This feature indicates the different reaction regimes between the reaction on $^{12}$C and $^{9}$Be, where  energy and angular momentum brought in by the collision differ considerably, as indicated in table~\ref{table1}. 
 
\begin{figure}[t]
\includegraphics[width=1\linewidth]{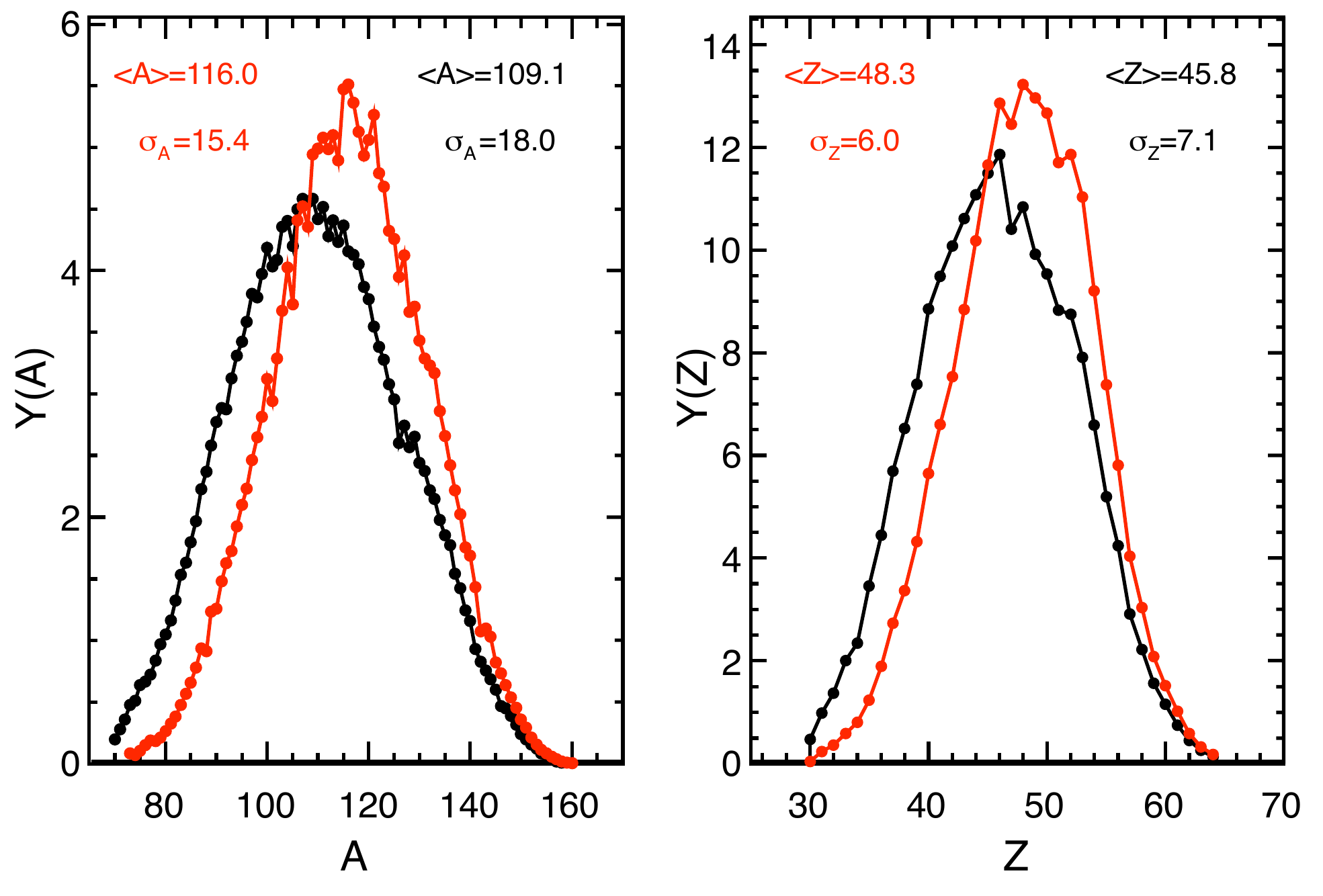}
 \caption{\label{A_Z} Mass (left) and atomic number (right) distributions of fission fragments produced in the two reactions $^{238}$U+$^{12}$C (black), and $^{238}$U+$^9$Be (red). The mean values as well as the width of the distributions are indicated with the same color code.}
 \end{figure}

The corresponding mass and atomic number distributions are shown in figure~\ref{A_Z} for both systems. The fission fragments produced in the $^{238}$U+$^9$Be reaction show a symmetric bell shape in mass and atomic number. The mean value of the fragment atomic-number distribution is 48.3 $\pm$ 0.4, which is coherent with the fission of the Cm compound nucleus, being the result of the complete fusion between $^{238}$U and $^9$Be. The confirmation of the atomic number of the compound nucleus shows that at this relatively high energy, the fusion is complete, and no significative proton evaporation arise after the formation of the compound nucleus. The mean value of the fragment mass distribution is 116 $\pm$ 0.8, indicating that in average 15 neutrons have been evaporated, including the post-scission emission of neutrons by the fragments. The fragment distribution is different for the  $^{238}$U+$^{12}$C reaction, where the atomic-number distribution shows an asymmetric pattern, with a mean atomic number of 45.8 $\pm$ 0.4, corresponding to a loss of six protons in average, with respect to the complete-fusion compound-system. The loss of protons may occur in incomplete fusion, pre-equilibrium emission, or proton evaporation from the compound nucleus. 
The widths of the mass and atomic number distributions are larger for the reaction induced on $^{12}$C target, reflecting the higher excitation energy of the fissioning system, resulting from the higher energy available in the centre of mass reference frame. 


\begin{figure}[t]
\includegraphics[width=1\linewidth]{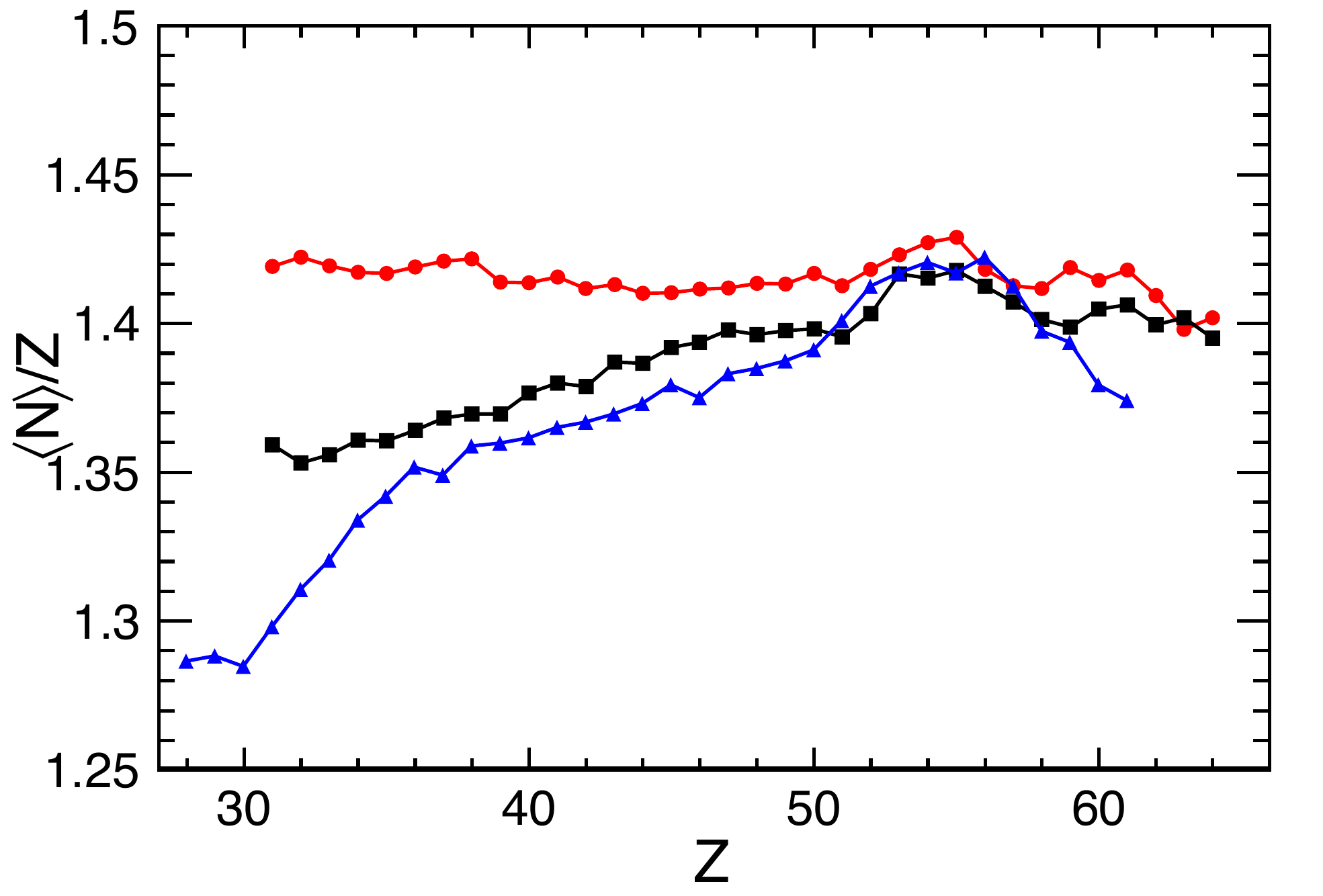}
 \caption{\label{NZ} Neutron excess of the fission fragments produced in the two reactions $^{238}$U+$^9$Be (red circles), and $^{238}$U+$^{12}$C (black squares). They are compared to similar data from a spallation reaction~\cite{PeB07} (blue triangles).}
 \end{figure}

In order to deepen the description of the gross properties of the fission-fragment distributions in both reactions,  the neutron excess is defined as the average neutron number $<N>$ of an isotopic distribution, divided by the corresponding atomic number $Z$. The evolution of the neutron excess of the fission-fragments with their atomic number measured in the two systems is displayed in figure~\ref{NZ}. In addition, the neutron excess of fission fragments produced in spallation reactions~\cite{PeB07} are displayed for comparison. 

\begin{figure}[t]
\includegraphics[width=1\linewidth]{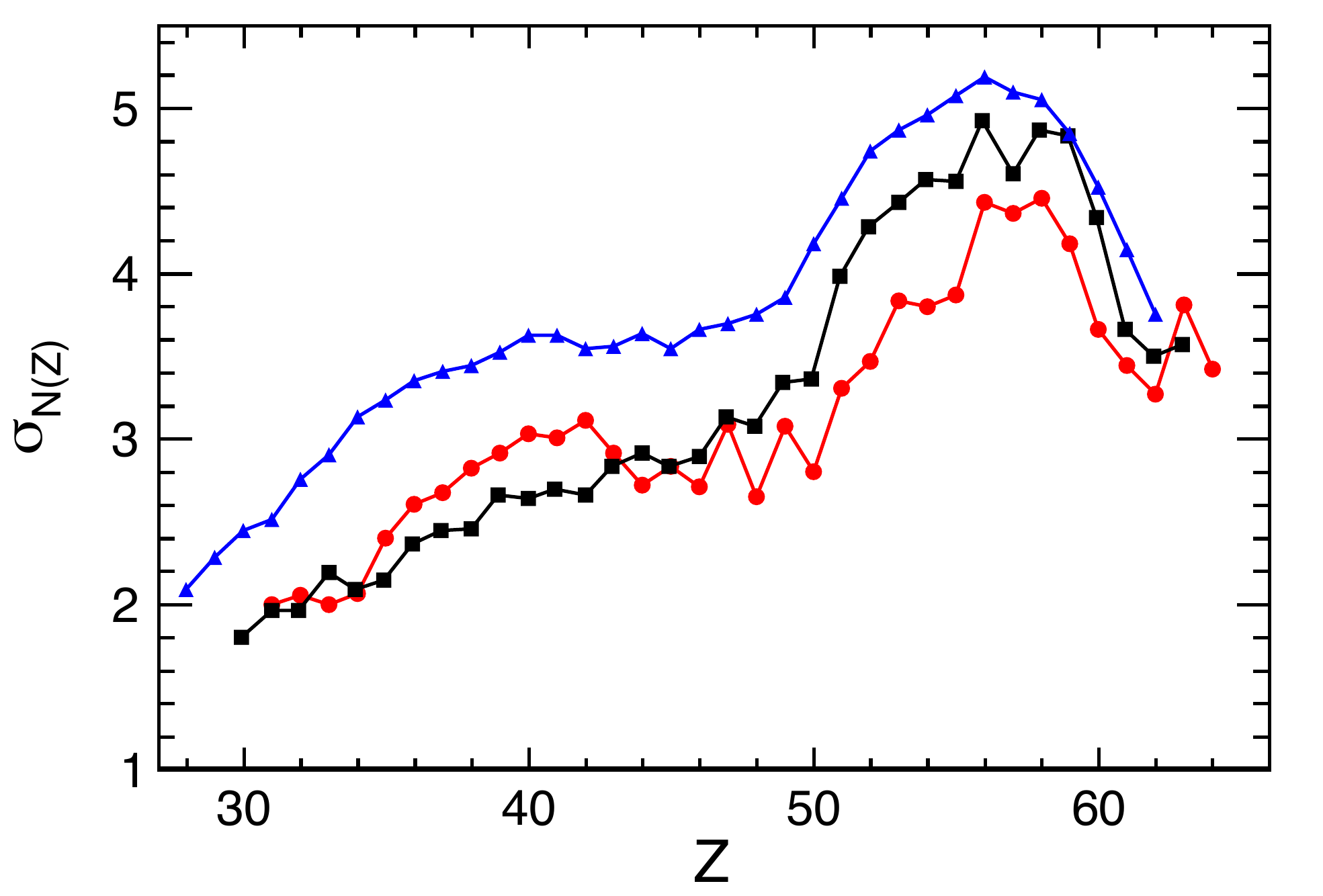}
 \caption{\label{SZ} Width of the isotopic fission fragments distributions in the two reactions: $^{238}$U+$^9$Be (red circles), and  $^{238}$U+$^{12}$C (black squares). They are compared to similar data from spallation reaction~\cite{PeB07} (blue triangles).}
 \end{figure}

The reaction $^{238}$U+$^9$Be at an average energy of 187 MeV in the centre of mass reference frame leads to the compound nucleus of $^{247}$Cm with an average excitation energy of 180 MeV, in case of complete fusion. The neutron excess of the fission fragments produced in this reaction shows a constant value of 1.42, over the complete element production. This average neutron- to proton-number ratio is compatible with the fissioning system $^{232}$Cm. The excitation energy of the compound nucleus is consistent with the evaporation of about 15 neutrons. The neutrons may be evaporated before the saddle point, during the elongation from saddle to scission, or after the scission point, depending on the time needed by the deformation process compared to the time for neutron evaporation.   The rather constant behavior of the fragments neutron excess for this system may lead to the conclusion that the compound nucleus is fissioning most probably at relatively high excitation energy, as almost no polarization induced by shell effects, known in low-energy fission~\cite{BoB89}, is observed with a significant amplitude. Still, a slight bump around $Z=54$ shows that a small component of low excitation-energy fission is present in the fragment distributions. Low-energy fission is influenced by nuclear shell-structure effects, and leads to the production of more neutron-rich fragments. The low-energy fission component is confirmed in figure~\ref{SZ}, where the width of the isotopic distributions is displayed as a function of the fragment atomic number. This parameter is linked to the temperature at which the nucleus fissions, as well as to the curvature of the potential energy at saddle~\cite{Arm70}. A steady and moderate increase of the isotopic-distribution width with the fragment atomic number is therefore expected~\cite{OgL85}. The sudden increase observed in the isotopic-distribution width in the vicinity of $Z=54$ confirms the presence of a second component of lower excitation energy fission, where shell effects induce the production of more neutron-rich fragments and increase significantly the width of the isotopic fission-fragment distributions expected from liquid-drop model considerations. In the case of the $^{238}$U+$^9$Be reaction, where the fusion is assumed to be complete from the fragment charge distribution (figure~\ref{A_Z}), the low-energy fission component appears after a long evaporation chain, which removes most of the excitation energy before fission. 

In the case of the $^{238}$U+$^{12}$C reaction, a similar but more pronounced bump is observed in the neutron excess of the fission fragments, as well as in the isotopic-distribution width. Both features indicate that the low-energy fission component is more important in this reaction. Different mechanisms may explain the contradictory observation that higher available excitation energy leads to a more significant contribution of low-excitation energy. Considering a complete fusion reaction, it could mean that the deformation time needed for the fused system to reach the saddle deformation is longer than at lower excitation energy, inferring the debated idea~\cite{HoI01} that the viscosity, {\it i.e.} the propensity that the nucleus has to transform single-particle excitation into collective excitation degrees of freedom, is  increasing with excitation energy. A possible influence of the angular momentum in this matter has been recently proposed~\cite{Ye11}. It is also likely that the low-enery fission component signs the dynamics of the entrance channel and the formation of the compound nucleus after incomplete fusion. This last assertion is supported by the similarity of the neutron excess and the isotopic-distribution width observed in the two reactions $^{238}$U+$^{12}$C and $^{238}$U+$^{2}$H at 1A GeV~\cite{PeB07}. In this spallation reaction, the collision stage of the reaction is known to produce a wide range of pre-fragments over a large range of excitation energy.  The resulting fission-fragment properties show a strong bump in the neutron excess as in the isotopic-distribution width that has been discussed in terms of the contribution to the fission yields of large impact-parameter collisions leading to low-excitation energy induced fission. An additional similarity between the $^{238}$U+$^{12}$C and the $^{238}$U+$^{2}$H reactions lays in the slope observed in the neutron excess of the light fragments, as displayed in figure~\ref{NZ}. This slope is stemming from the different fissioning systems that contribute to the fission-fragment distributions. The lighter fragments are produced mostly in the fission of lighter fissioning systems, which are produced with lighter neutron excess, resulting either from incomplete fusion or pre-equilibrium emission in the case of the $^{238}$U+$^{12}$C reaction, or from long intra-nuclear nucleon-nucleon collisions cascade in the case of   spallation  reaction. The neutron-excess of the fission-fragments produced in the $^{238}$U+$^9$Be reaction shows a constant trend, confirming that fission fragments originate from similar fissioning systems, inferring a complete-fusion process. A more detailed description of the measured observables would require a complete simulation of the nuclear reactions at these energies, including the entrance channel description and its statistical decay, that goes beyond the scope of the present work.  

  
\section{Conclusions}

Inverse kinematics coupled to the use of a high-resolution spectrometer is shown to be a powerful tool to identify and measure the isotopic yields of fission fragments produced in the collision of a $^{238}$U beam at 24 MeV/u  with $^9$Be and $^{12}$C targets. The isotopic yields measured in these two reactions show that the production of neutron-rich isotopes is favoured in the reaction where complete fusion at lower excitation energy arises. Isotopic distribution over the complete fission-fragment production is a measurement of new kind in the field of fission dynamics, and new properties of the fission fragment distributions such as the neutron excess are extracted. These properties allow to investigate deeper the fusion-fission mechanism than the mass distributions that were measured up to now~\cite{YaH05, ItI10}. The comparison of the atomic-number and mass distributions combined with the analysis of the  isotopic-distributions properties show that between the $^9$Be and the $^{12}$C target the reaction changes substantially of regime, evolving from a complete fusion reaction to incomplete fusion.

\begin{acknowledgements}
The authors acknowledge the GANIL accelerator staff for providing the $^{238}$U beam. O. D. acknowledges the hospitality of Michigan State University. 
\end{acknowledgements}

\bibliography{FF.bib}
\end{document}